\documentclass{article}

\pdfoutput=1

\usepackage[T1]{fontenc}
\usepackage[utf8]{inputenc}
\usepackage[english]{babel}
\usepackage{indentfirst}

\usepackage{graphicx}
\usepackage{subfig}
\usepackage{float}
\usepackage{placeins}
\usepackage{caption}
\usepackage{tikz}

\usepackage{booktabs}
\usepackage{multicol}
\usepackage{multirow}

\usepackage{amsmath}
\usepackage{amssymb}
\usepackage{mhchem}
\usepackage{siunitx}
\usepackage{physics}

\usepackage{url}
\usepackage{bbold}
\usepackage{cleveref}
\usepackage{authblk}
\usepackage[a4paper, margin=30mm]{geometry}
\usepackage[backend=biber, citestyle=numeric, sorting=none]{biblatex}
\title{Prospects for single photon sideband cooling\\in fermionic Lithium}

\author[1]{\small F. Berto}
\author[2,3]{\small C. Sias}

\affil[1]{\small Politecnico di Torino, Corso Duca degli Abruzzi, 24, 10129 Torino, Italy}
\affil[2]{\small Istituto Nazionale di Ricerca Metrologica, Strada delle Cacce, 91, 10135 Torino, Italy}
\affil[3]{\small European Laboratory for Non-Linear Spectroscopy, Via Nello Carrara, 1, 50019 Sesto Fiorentino, Italy}

\date{\small \today}

\begin{document}
\maketitle

\begin{abstract}
We present an analytic and numerical study for realizing single photon sideband cooling in an ultracold sample of fermionic Lithium trapped in a periodic optical potential. We develop an analytical model and obtain a master equation for the bound level populations. The cooling sequence is simulated both with a laser at a fixed frequency and with a frequency sweep. Finally, a Monte Carlo simulation is performed taking into account the full hyperfine spectrum of \ce{^6Li}. We find that a gas of \ce{^6Li} atoms loaded from a Magneto-Optical trap into a deep optical lattice can be cooled down to a \SI{99}{\%} occupancy of the lattice ground state after a \SI{5}{\milli\second} single photon sideband cooling using the D1 line of Li. 
\end{abstract}

\section*{Introduction}
Resolved sideband cooling (RSC) is a laser cooling technique used to cool trapped particles to their ground state of motion. RSC has been succesfully implemented in several experiments involving trapped ions\cite{Diedrich1989} and ultracold atoms loaded in an optical lattice\cite{Hamann1998}.

In order to introduce the working principle of RSC, let us consider the particle as a two-level system (TLS), with states \( \ket{g} \), \( \ket{e} \) and natural linewidth of the excited state \( \gamma \). Let us also assume that the ground and the excited states perceive the same trapping potential and that the trapping potential is harmonic. In this model each state of the TLS is transformed in a manifold \( \ket{i, n}\), where \( i = g, e \) and where \( n \) represents the quantum number of the motional state in the harmonic oscillator. RSC is a two-step process. In a first step the motional state of a trapped particle is changed by driving the \( \ket{g, n}\to\ket{e, m} \) transition with a laser set at a frequency \( \omega_L = \omega_0 + (m-n)\omega_T \), where \( \hbar\omega_0 \) is the energy difference between ground and excited states in the TLS, while \( \omega_T \) is the harmonic trap frequency. In the second step the particle internal state goes back to the \( \ket{g} \) internal level, \emph{e.g.} by spontaneous emission, but to a lower motional state with respect to the initial one. If the states of the TLS are long-lived states this second step of RSC is usually realized through optical pumping, as in Raman sideband cooling.

In order to be efficient, sideband cooling requires two conditions to be satisfied.\\
The first condition is that the motional level separation must be larger than the recoil energy associated with the laser absorption. This is quantified by the \emph{Lamb-Dicke parameter} \( \eta = k_z z_0 \), where \( k_z \) is the projection of the laser wavevector along the harmonic oscillator axis and \( z_0 \) is the spread of the zero-point harmonic wavefunction.\\
For implementing an efficient sideband cooling, the system must be in the \emph{Lamb-Dicke regime}, \emph{i.e.} \( \eta \ll 1 \). In this regime it is possible to expand the light-atom interaction operator in powers of \( \eta \)\cite{Javanainen1981}: the terms of the series expansion represent sideband transitions associated with the loss or gain of one or more motional quanta. The transition that leaves the motional state unchanged (\( \Delta n = 0 \)) is referred to as the \emph{carrier transition}, while the \emph{blue} and \emph{red sidebands} are the transitions associated with the gain or loss of a vibrational quanta (\( \Delta n = \pm1 \)), respectively.\\
The second requirement is for the system to be in the so-called \emph{strong coupling condition}, \emph{i.e.} \( \nu_\text{trap} > \gamma \), which ensures that sidebands and carrier are well resolved spectroscopically, so that they can be separately addressed by using laser light\cite{Stenholm1986}. Since the carrier transition is the most intense, spontaneous emission will occur mainly without changing the vibrational number. Therefore, it is possible to cool the system through the optical cycle: \( \ket{g, n} \to \ket{e, n-1} \to \ket{g, n-1} \).

Usually, single photon sideband cooling is performed on trapped ions since deep trapping potentials can be conveniently created with electric fields, and narrow optical transitions are available in most of the elements that are used for ion trapping. In cold atoms experiments, instead, narrow optical transitions are often not at hand --- with the exception of alkaline-earth and alkaline-earth-like atoms. Therefore, in order to resolve the sidebands, 2-photon Raman processes between hyperfine levels are usually employed (see Ref.\cite{Parsons2015} for an example of a recent work on Lithium).

In this work we investigate the possibility of performing single photon sideband cooling on the D1 transition of neutral fermionic Lithium atoms trapped in a periodic optical potential. Single photon sideband cooling has two main benefits with respect to Raman transitions: first, the experimental setup is simpler since only one laser is required for cooling; second, the transition rate is higher and the cooling process in principle faster. In our investigation, we start from a TLS in a non-harmonic trap with different potentials for ground and excited states, and study RSC by using an analytical and a kinetic Monte Carlo simulation\cite{Kratzer2009}. We then extend our simulation to include the hyperfine levels (HFS) of Lithium atoms at low intensity magnetic fields.

\section*{Mathematical model}
We consider a relatively high-finesse (\( f \approx 17000 \)) cavity resonating at \SI{903.4}{\nano\metre} for trapping \ce{^6Li} atoms. The resulting dipole potential can be made deep enough that the system satisfies the strong coupling condition and is in the Lamb-Dicke regime. If we consider a cavity input power of \SI{100}{\milli\watt}, the \( \ce{^2S_{\frac{1}{2}}} \) and \( \ce{^2P_{\frac{1}{2}}} \) levels experience a trapping potential of depth \SI{26}{\milli\kelvin} and \SI{22}{\milli\kelvin}, respectively. The equivalent harmonic trapping frequencies are \( \nu_g = \SI{9.5}{\mega\hertz} \) and \( \nu_e = \SI{8.7}{\mega\hertz} \). However, the harmonic approximation works only for the lowest motional energy levels, and in the calculations reported in this work we considered for the energy of the bound levels the mean energy of the corresponding Bloch bands.

\begin{figure}
	\centering
	\subfloat[][Carrier and sidebands transitions spectrum for different starting values of the motional quantum level. The x-axis reports the detuning with respect to the unperturbed TLS transition. For display purposes the carrier height has been reduced by a factor of 100.\label{fig:sb_detuning}]
		{\includegraphics[width=0.45\columnwidth]{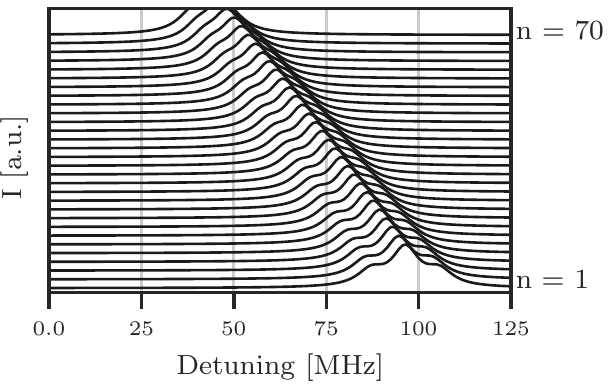}}~
	\subfloat[][Hyperfine structure of \ce{^6Li}. The three arrows indicate the transitions that must be simultaneously addressed for efficient sideband cooling.\label{fig:hfs_structure}]
		{\includegraphics[width=0.45\columnwidth]{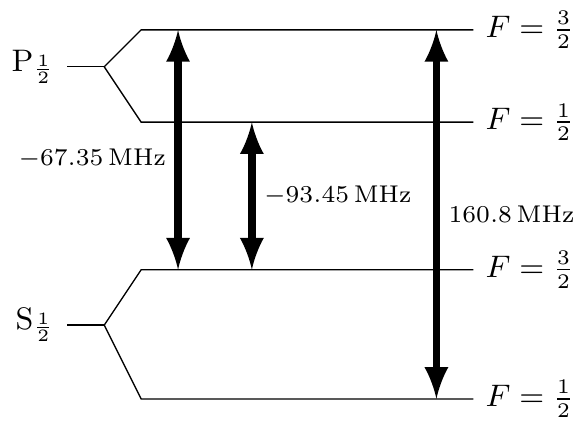}}
	\caption{}
\end{figure}

In such conditions the system has approximately 70 bound levels, although the actual number of motional levels is slightly different for the ground and the excited internal states. Therefore, there is a number of bound levels in the ground state potential that cannot be cooled via RSC because there is no counterpart in the exited state energy spectrum. Both the strong coupling and the Lamb-Dicke regime conditions are satisfied for \( 80\% \) of the bound energy levels\footnote{The remaining levels are the least bound ones. They have a sideband-carrier separation similar or smaller than the natural linewidth.}. Since the separation between the levels is not constant, carrier and sidebands associated with a different starting level do not coincide in frequency but rather spread out over a wide frequency range. \Cref{fig:sb_detuning} shows the excitation spectrum for different initial bound states, where the front-most curve represents the spectrum of a particle in the lowest motional state \( n = 0 \).

The population of the different energy levels can be calculated analytically by building a master equation. To this end, we modify the model of Ref.s\cite{Javanainen1981, Stenholm1986} to include the dependence of the energy spectrum on the motional quantum number \( n \), and find:

\begin{equation}
	\label{eq:master}
	\begin{split}
	\dot{P}(n,t) &= (n+1)\ A_-(n+1)\ P(n+1) \\
							&\quad + n\ A_+(n-1)\ P(n-1) \\
							&\quad - [(n+1)A_+(n)\ + n A_-(n)]\ P(n)\,,
	\end{split}
\end{equation}
where \( A_{\pm} = \eta^2 \qty[\alpha \Gamma(\nu) + \Gamma(\nu - \nu_{n\to n\pm1})] \) are the heating and cooling rates, \( \nu_{n\to n\pm1} \) is the difference in energy between the \( n \)-th and \( (n\pm1) \)-th motional levels divided by the Planck constant, \( \Gamma(\nu) \) is the scattering rate of the TLS for a laser of frequency \( \nu \), \( P(n) \) is the population of the n-th vibrational level, and \( \alpha = \flatfrac{2}{5} \) is an average angular factor in spontaneous emission.

One must also take into account the fact that the bound states \( \ket{n}_g \) and \( \ket{m}_e \), corresponding to motional levels of the ground and excited states, respectively, are not orthogonal, since they are associated to different trapping potentials. This non-orthogonality leads to an additional cooling and heating rate of the zero-th order in the Lamb-Dicke parameter\cite{Taieb1994}. We account for this effect by including the terms

\begin{equation*}
	\label{eq:zeroth_terms}
	R_{n\to m} = \abs{\braket{e, m}{g, n}}^2 \ \qty[\alpha \Gamma(\nu) + \Gamma(\nu - \nu_{n\to m})]
\end{equation*}
in \cref{eq:master}. We approximate the value of the coefficients \( \braket{e, m}{g, n} \) by calculating the overlap integral between harmonic oscillator wavefunctions (an analytical expression can be found in Ref.\cite{Babusci2011}). Finally, we add a \( A_\pm^{\text{trap}} \) in \cref{eq:master} in order to consider the off-resonant scattering of light from the trapping laser, and we estimate a scattering rate of 5 photons per second.

In order to calculate the steady-state population of the different motional levels, we must solve the differential equation:
\begin{equation*}
	\dot{\va{P}} = \mathbf{A}\va{P}\,.
\end{equation*}
The population vector \( \va{P} \) is defined by \( {\{\va{P}\}}_n = P(n, t) \) and \( \mathbf{A} \) is the coefficients matrix of the modified master equation, so that \( {\{\mathbf{A}\}}_{n,m} \) represents the rate of transition between the states \( \ket{n}_g \) and \( \ket{m}_g \).\\
If the matrix \( \mathbf{A} \) is time-independent\footnote{The requirement that \( A_\pm(t) = A_\pm \) in turn requires that the scattering rate \( \Gamma(\nu) \) is constant, \emph{i.e.} that the laser frequency is fixed.} a solution is given by
\begin{equation}
	\label{eq:linalg_sol}
	\va{P} = \sum_i c_i \va{v}_i e^{\lambda_i t}\,,
\end{equation}
where \( \va{v}_i, \lambda_i \) are the eigenvectors and eigenvalues of the coefficient matrix, and \( c_i \) are integration constants.

The solution of \cref{eq:linalg_sol} shows that the cooling with a laser at a fixed frequency is inefficient. Qualitatively, this is due to the fact that a laser resonant to the red sideband of the n-th level will scatter mainly photons in the blue sideband (and thus cause heating) of the k-th levels with \( k > n \) (see \cref{fig:sb_detuning}). To solve this problem, we introduce in our model a frequency sweep of the sideband cooling laser. The laser light is initially tuned to the red sideband of a high energy bound level, and the frequency is then swept toward larger frequencies, in order to favour the cooling of lower energy levels. In this way the laser scans all the red sidebands from the most excited levels to the ground state, thus favouring the cooling of the largest possible number of atoms. In order to calculate the optimal frequency sweep for the cooling laser, we cannot use \cref{eq:linalg_sol}, since the cooling and heating rates \( A_\pm \) can not longer be considered constant. We solve this issue by discretizing time in time lags (of duration \( \dd{t} = \SI{1}{\micro\second} \)) during which we assume \( A_\pm \) to be constant. In this way, we solve \cref{eq:linalg_sol} iteratively over each time interval.

In order to calculate the initial distribution of the atoms in the optical potential's bound states, we consider that the atoms are suddenly transferred from a magneto-optical trap (MOT) to the periodic potential. We assume that a single atom in the MOT is well approximated by a Gaussian wavepacket

\begin{equation*}
	\psi_\sigma(x) = \frac{1}{\sqrt{\sigma\sqrt{2\pi}}}\, e^{-\frac{x^2}{4\sigma^2}}\, e^{i\frac{p_0}{\hbar} x}
\end{equation*}
where \( \sigma \) is the spatial extension of the wavepacket at \( t=0 \) and \( p_0 \) is the momentum associated with the group velocity of the wavepacket. We assume \( \sigma = \SI{500}{\micro\metre} \) and \( p_0 = \sqrt{m k_B T_0} \), with \( k_B \) the Boltzmann constant and \( T_0 = \SI{40}{\micro\kelvin} \) the initial temperature of the atoms\cite{Burchianti2014}. The population of the n-th bound level is \( P_n = c_n^2 \), where: 

\begin{equation*}
	c_n = \frac{1}{\sqrt{\sigma\sqrt{2\pi}}}\, \frac{1}{\sqrt{2^n n! \sqrt{\pi}}}\, \qty(\frac{m \omega_T}{\hbar})^{1/4} \int_{-\infty}^{\infty} e^{-\qty(\frac{m \omega_T}{2\hbar} + \frac{1}{4\sigma^2}) x^2\, +\, i\frac{p_0}{\hbar} x}\, H_n\qty(\sqrt{\frac{m \omega_T}{\hbar}}\, x)\
\end{equation*}
By substituting the Hermite polynomials \( H_n(x) \) with the Gould-Hopper polynomials\footnote{We note that these are the solutions to the heat Fourier equation and that they are different from the usual definition of bivariate Hermite polynomials.} \( H_n(x,y) \) using the relation \( H_n(2x, -1) = H_n(x) \), the integral can be rewritten in a solvable form\cite{Babusci2011}. We calculate for our starting condition an initial average occupation number of \num{23}.

All calculations and simulations were performed by using a code written in  Python and the open-source Scipy ecosystem\cite{Scipy2019}. The characteristic values of the Mathieu equation were also calculated numerically\cite{Coisson2009}.

\section*{Results}
\begin{figure}
	\centering
	\subfloat[][Average occupation number of the motional energy levels in the two-level system approximation. The solid line represents the result obtained by simulating a cooling experiment with a fixed laser frequency, while the dashed line represents the result for a laser frequency sweep.\label{fig:cooling_TLS_result}]
		{\includegraphics[width=0.49\columnwidth]{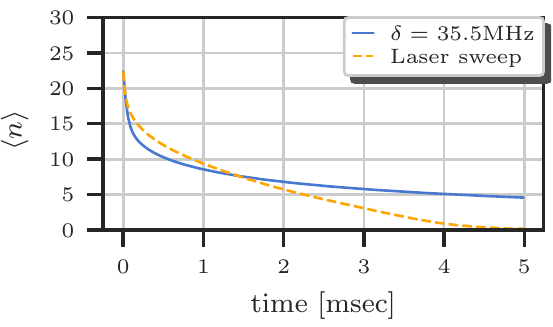}}~
	\subfloat[][Average occupation number of motional energy levels considering all 12 hyperfine levels of the ground state of \ce{^6Li}. The solid line represents the average occupation number of the different hyperfine levels, the the dashed lines the average level population for each HF state.\label{fig:cooling_full_result}]
		{\includegraphics[width=0.49\columnwidth]{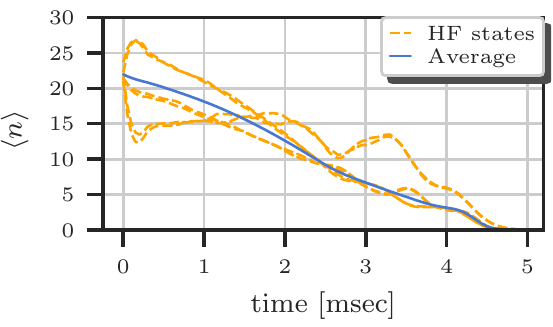}}
	\caption{}
\end{figure}

\Cref{fig:cooling_TLS_result} shows the average energy level as a function of the time in a simulated \SI{5}{\milli\second} cooling in the TLS approximation. We first simulate a cooling sequence with a fixed frequency laser and found that the initial average occupational number can be reduced by \( \Delta n_{\%} = \flatfrac{(n_0 - n_{\text{fin}})}{n_0} = \SI{79.6}{\%} \), where \( n_0 \) and \( n_{\text{fin}} \) are the initial and final average occupation numbers, respectively, at an optimal detuning of \( \delta = \SI{35.5}{\mega\hertz} \)\footnote{The detuning is relative to the unperturbed Bohr frequency of the TLS\@. This value is obtained by simulating many experiments with different laser frequencies and then choosing the frequency that maximizes \( \Delta n_{\%} \).}. This result slowly improves for times longer that \SI{5}{\milli\second}, but only by a few percent. A considerably more efficient cooling is observed when using a laser sweep from \( \delta_0 = \SI{26.2}{\mega\hertz} \) to \( \delta_1 = \SI{82.9}{\mega\hertz} \): \( \Delta n_{\%} =\SI{99.3}{\%} \).

In order to include the HFS, one has to take into account the partial overlap between sidebands of different hyperfine levels of \ce{^6Li}. This may lead to the excitation of blue sidebands for some of the HFS\@. This effect can be compensated by using more than one laser frequency at the same time (see \cref{fig:hfs_structure}). One may still write a master equation for the full system, but the analytical solution would be affected by the round-off error arising in the diagonalization of a large coefficient matrix. Instead, we performed a kinetic Monte Carlo simulation of the master rate equation. The three required frequencies are swept together between the same detunings \( \delta_0 \) and \( \delta_1 \)\footnote{Detunings from the corresponding unperturbed hyperfine transition.}. \Cref{fig:cooling_full_result} shows the result of the simulation, the motional number is reduced (in average over all HFS levels) by \( \Delta n_{\%} =\SI{99.9}{\%} \). All the calculations are valid for low homogeneous magnetic fields, up to 10 Gauss. For higher magnetic fields we observe that cooling is no longer efficient for some of the HF states, in particular when the magnetic field is increased over 20 Gauss.

The separation between red sidebands of two consecutive motional levels is on the order of \SI{1}{\mega\hertz}. It is therefore important to stabilize the trap depth in order to avoid a shift of the motional levels spectrum. We calculate that in order to maintain cooling efficiency above \( \SI{95}{\%} \) the laser power should be stabilized to a maximum \SI{5}{\milli\watt} fluctuation, \emph{i.e.} \( \flatfrac{\Delta P}{P_0} = \SI{5}{\%} \), a value that can be achieved with a conventional stabilization of the laser power.

\FloatBarrier

\section*{Acknowledgements}
We thank M. Zaccanti, L. Duca, E. Perego for helpful discussions.\\
This work was financially supported by the ERC Starting Grant PlusOne (Grant Agreement No. 639242), and the FARE-MIUR grant UltraCrystals (Grant No. R165JHRWR3).

\printbibliography{}

\end{document}